\documentclass[aip,reprint,amsmath,amssymb]{revtex4-1}

\usepackage{graphicx}
\usepackage{dcolumn}
\usepackage{bm}

\usepackage[utf8]{inputenc}
\usepackage[T1]{fontenc}
\usepackage{mathptmx}
\usepackage{xr}
\draft 

\begin{document}




\preprint{AIP/123-QED}

\title[snapshot attractors in SIR]{How can contemporary climate research help to understand epidemic dynamics? -- Ensemble approach and snapshot attractors}


\author{T. Kov\'acs}
 \email{tkovacs@general.elte.hu}
\affiliation{ 
Institute of Theoretical Physics, E\"otv\"os University, P\'azm\'any P. s. 1A, H-1117 Budapest, Hungary.
}%


\date{\today}

\begin{abstract}
Standard epidemic models based on compartmental differential equations are investigated under continuous parameter change as external forcing. We show that seasonal modulation of the contact parameter superimposed a monotonic decay needs a different description than that of the standard chaotic dynamics. The concept of snapshot attractors and their natural probability distribution has been adopted from the field of the latest climate-change-research to show the importance of transient effect and ensemble interpretation of disease spread. After presenting the extended bifurcation diagram of measles, the temporal change of the phase space structure is investigated. By defining statistical measures over the ensemble, we can interpret the internal variability of the epidemic as the onset of complex dynamics even for those values of contact parameter where regular behavior is expected. We argue that anomalous outbreaks of infectious class cannot die out until transient chaos is presented for various parameters. More important, that this fact becomes visible by using of ensemble approach rather than single trajectory representation. These findings are applicable generally in nonlinear dynamical systems such as standard epidemic models regardless of parameter values.
\end{abstract}

\pacs{SIR model, seasonal forcing, ensembles, snapshot attractor, transient chaos}

\maketitle 

\section{\label{sec:background}Background}

The effort to estimate short and long-term behavior, system parameters, possible control strategies, and also social impact of COVID-19 pandemic stimulated an explosion of disease modelling\footnote{See the pre-filtered search on http://arxiv.org and https://connect.biorxiv.org/relate/content/181} and scientifically demanding examinations from many different point of view made by various community members and research groups around the globe. Although the present situation is rather new, we have many examples from the history that helped to lay down basic mathematical rules of disease propagation. The celebrated SIR (susceptible - infectious - recovered) model\cite{Kermack1927} splits the population into three disjoint parts and deals with the number of individuals in these sub-populations as time goes on.

The SIR model and its variants [SI, SIS, SEIR, SEIRS, RAS] show qualitatively similar dynamics and are in good agreement with observations. In a homogeneous environment these models possess a globally stable fixed point attractor as a disease equilibrium \cite{Keeling2001}. Omitting the stochastic nature of real world disease spread  \cite{Tuckwell2007,Allen2008,Olsen1990,Witbooi2013}, the deterministic SIR-like models present diverse and reach dynamics. The non-linearity of the model \cite{Olsen1988,Rand1991,Bolker1993,Keeling1997} and also the time dependent internal forcing can be considered as a source of complexity.  

Sufficiently large seasonal forcing or different mixing rates between sub-populations can cause large oscillations and also period doubling cascades \cite{London1973,Aron1984,Schwartz1985,AndersonBook,Altizer2006,Stone2007,Rozhnova2009}. It has also been demonstrated that for certain values of system parameters the non-autonomous models of recurrent epidemics (measles, mumps, rubella, H5N1 avian influenza) show chaotic behavior \cite{Olsen1990,ORegan2013,Barrientos2017,Duarte2019}. Traditional SIR-like epidemic models are dissipative nonlinear low-dimensional systems with constant, periodic, quasi-periodic\cite{Bilal2016}, or term-time\cite{Keeling2001,Papst2019} external forcing whose dynamics is governed by (chaotic) attractors in phase space. Furthermore, even if the long-term dynamics is regular and the final state of the system is a fixed-point attractor the route to this condition might be rather complex. Many studies examined the role of finite-time irregularity in ecological models\cite{Hastings1994,Hastings2004,Singh2011,Morozov2020} as well as in epidemic dynamics\cite{Rand1991,Earn2000,Hempel2015,Papst2019} concluding the relevance of transient behavior. 

It is known from dynamical systems theory that chaotic attractors always can be characterized by the natural measure corresponding to the distribution of possible states in phase space \cite{Eckmann1985,OttBook,GruizBook}. It turned out, however, that traditional numerical methods, such as monitoring a single chaotic long-term trajectory, fail in case of arbitrary forcing (like smoothly varying parameters). Obviously, if one wants to model such a system (e.g. the changing climate with shifting atmospheric CO$_2$ concentration), this shortcoming has to be overcome. The mathematical concept of \textit{snapshot attractors}\footnote{For completeness, it should be mentioned that a generalization of snapshot attractors has been done and is referred to as pullback attractors in mathematical and climate research. \cite{Ghil2008,Chekorun2011}}, known for many years in (theoretical and experimental) dynamical systems community  \cite{Romeiras1990,ArnoldBook,Lai1996,Jacobs1997,Neufeld1998,Karolyi2004,Sommerer1993,Vincze2017}, fulfills entirely our wish.

Briefly, snapshot attractors are time-dependent objects in phase space of non-autonomous dynamical systems. Thus the shape of snapshot attractors is changing in time while their fractal dimension might even remain constant\cite{Karolyi2004}. Furthermore, obtaining one single trajectory in a system with arbitrary driving force, does not provide the same result as the ensemble approach (many trajectories emerging from slightly different initial conditions) in the same system at a given time instant. This effect is the consequence of the fact that ergodicity is not satisfied as the system is driven aperiodically \cite{Drotos2016}. This conclusion generated the recent opinion that the changing climate should be scrutinized by ensemble approach (\textit{parallel climate realizations}) rather than averaging a single long-term time series \cite{Bodai2012,Daron2013,Drotos2015,Herein2016,Pierini2016,Tel2019}. 

In this work a seasonally forced deterministic epidemic model with monotonically changing contact rate (due to, for instance, vaccination or restricting the movement of the population) is presented. We propose that the statement by Ref.~\onlinecite{Bodai2012} \textit{''climate change can be seen as the evolution of snapshot attractors''} also holds for disease spread dynamics with continuously changing contact parameter.

In Section~\ref{sec:model} the epidemic model is defined. After that, the results about stationary (Sec.~\ref{sec:stat_epidemic}) and changing epidemic (Sec.~\ref{sec:changing_epidemic}) are presented. Section~\ref{sec:concl} is devoted to conclusions.

\section{\label{sec:model}Standard epidemic model}

Compartmental disease models describe the number of individuals in a population regarding their disease status: susceptible ($S$), infectious ($I$), or recovered ($R$). Although, these models involve many simplifications (such as the progression of infection or  difference in response of individuals) they performed well in real world epidemic situations \cite{Keeling2005}. There are two major groups of epidemic models: the SIR-like cluster that characterizes lifelong immunity (e.g. measles, whooping cough), and the SIS-like class (containing mostly sexually transmitted diseases) which portrays repeated infections\cite{Keeling2005}.         

Here we study the SEIR equations \cite{Anderson1982} that involves a fourth group in addition to previous ones. More concretely, we assume that an individual enters the population at birth as susceptible and leaves it by death. A susceptible becomes exposed ($E$) when contacting one or more persons, called  infective(s), that can transmit the disease. In an incubation period the exposed individuals are infectious but are not yet infectious. After this term they become infective and become later immune or recovered. 

The mathematical model associated with above description reads as follows
\begin{equation}
\begin{split}
\frac{\mathrm{d}S}{\mathrm{d}t}& = m-bSI-mS, \\
\frac{\mathrm{d}E}{\mathrm{d}t}& = bSI - (m+a)E, \\
\frac{\mathrm{d}I}{\mathrm{d}t}& = aE- (m+g)I, \\
\frac{\mathrm{d}R}{\mathrm{d}t}& = gI - mR
\end{split}
\label{eq:seir}
\end{equation}
with the following notations and assumptions \cite{Dietz1975}: (i) $S,\,E,\,I$ and $R$ are smooth functions of time and the size of the whole population remains unchanged, $S+E+I+R=1;$ (ii) there are equal birth and death rates ($m$); (iii) the probability that an exposed individual remains in this class for a time period $\tau$ after the first contact is $\exp(-a\tau),$ where $1/a$ is the mean (or characteristic) latent period; (iv) similarly $1/g$ gives the mean infectious period, the time period that an individual spends as infectious before recovery; (v) immunity is permanent and recovered individuals do not re-enter to susceptible class. The contact rate $b(t),$ the average number of susceptibles contacted per a single infective per unit time, is the origin of the spread of the disease. In the case of annual periodicity
\begin{equation}
    b(t) = b_0(t)[1 + b_1\cos(2\pi t)].
    \label{eq:bt}
\end{equation}
and $t$ is measured in units of years.

When the system parameters are kept constant ($m,\,a,\,g,\,b_0,\,b_1=0$), the solution of Eq.~(\ref{eq:seir}) shows weakly dumping oscillation with a globally stable equilibrium ($b>g$). In this case the dynamics can be further characterized by the expected number of secondary cases caused by an infectious individual in the susceptible population. This is known as the basic reproductive ratio, and can be expressed  denoted here by $R_0=ba/[(m+a)(m+g)].$\cite{Aron1984} $R_0$ essentially determines whether a disease can ($R_0>1,$ endemic equilibrium) or cannot ($R_0<1,$ infection-free) persist in a population. Provided that latent and infectious periods are short, i.e. $a,g\gg m,$ one can use the approximation $R_0=b/g.$

There are observations, for example in childhood diseases or avian influenza, that do not show damped characteristics rather (ir)regular cycles instead. This phenomenon can be linked to seasonal variation in contact rate $b(t)$ or in recruitment rate $g$ \cite{ORegan2013}. It has already been noticed that in case of periodic forcing, $b_0(t)=b_0=$const. in Eq.~(\ref{eq:bt}), chaos emerges in epidemic time series \cite{Keeling2001,Katriel2012,Duarte2019}. Similarly to the Lorentz84 model \cite{Lorenz1984} wherein the solar irradiation induces seasonal effect, the SEIR model with variable contact rate is a non-autonomous low-order system with well-defined chaotic attractor in the phase space at certain parameter values. The low dimension of the phase space allows to visualize its pattern comfortably.

As a result of the fixed population size, point (i) above, one of the four equations (traditionally the equation for $R$) can be omitted. In addition, the infectious and exposed groups turn out to be linearly related to the first order \cite{Aron1984} at least, see Figure~S1 in supplementary material. Consequently, the $S-I$ phase portrait represents the state space texture accurately. Plotting one trajectory in the $S-I$ plane of the phase space one would observe a coil design due to the explicit time dependence of the model, $b_1\neq 0$. A conventional method to make a periodically forced system \textit{autonomous} is to take ''pictures'' about the phase space with the same frequency as the excitation acts, i.e. making a \textit{stroboscopic map} \cite{OttBook,GruizBook}. By choosing an initial time $t_0$ and purely sinusoidal force in Eq.~\ref{eq:bt} ($b_0(t)=$const.) one can interpret the filamentary shape of the chaotic attractor by defining the stroboscopic map $t_0 \mod T = 0.$ Therefore, the periodic variation in Eq.~(\ref{eq:bt}) yields am \textit{annually stationary epidemic} with steady attractor.

Eq.~(\ref{eq:bt}) assumes a constant average contact rate $b_0.$ We have an annual period starting with its maximum at $t=0.$\footnote{Choosing a different initial phase does not effect the qualitative picture of the dynamics. Only the shape of the phase space object differ to those obtained in present study.} In physical or biological context this means the largest contact rate possibly due to school term, vacation and holidays, seasonal breeding pattern in a seabird colony, etc. Eq.~(\ref{eq:seir}) with the following parameters, corresponding to measles\cite{Olsen1990}, shows irregular dynamics: $m=0.02$ year$^{-1}$, $a = 35.84$ year$^{-1}$, $g = 100$ year$^{-1}$, $b_0=1800$ year$^{-1}$, $b_1 = 0.28$ year$^{-1}$. 
The above parameters lead to $R_0\approx 18.$ Numerical calculations show that for lower mean values of the contact rate periodic fixed-point attractors exist. For instance, biannual cycles at $b_0\approx1500.$ For more detailed asymptotic dynamics see the bifurcation diagram in Figure~\ref{fig:bif_diagram} (blue curve).  

Until now $b_0$ was kept as a constant. Following the climate change methodology monotonic variation in $b(t)$ through $b_0(t)$ is established. Equation~(\ref{eq:b0t}) specifies the temporal decay of the mean contact rate 
\begin{equation}
    b_0(t)=
\begin{cases}
1800 &\text{if } t\leq t_{\mathrm{st}}\\
1710e^{-\alpha (t-t_{\mathrm{st})}}+90 &\text{if } t > t_{\mathrm{st}}.
\end{cases}
 \label{eq:b0t}
\end{equation}
Here $t_{\mathrm{st}}$ represents the time until the epidemic is stationary, i.e. $b_0=$ const., this is set to be 250 years after initialization. This amount of time seems to be enough since the convergence of trajectories to the attractors is much shorter, $t_{\mathrm{c}}\approx 50$ years. The exponent $\alpha$ is the decay rate of $b_0(t).$ In this study $\alpha$ takes three values 0,04 yr$^{-1},$ 0.01 yr$^{-1},$ and 0.004 yr$^{-1},$ and $b_0$ always starts at 1800, i.e. from the chaotic regime. Eq.~(\ref{eq:b0t}) implies that $\lim_{t\to \infty} b_0(t)=90.$ This means that the critical value of mean contact rate $b_0=100$ (or in terms of basic reproduction rate $R_0\approx 1$) is reached at different times after $t_{\mathrm{st}}$, according to $\alpha.$ 

The main motivation of this study is based on Eq.~(\ref{eq:b0t}). Considering, for example, the time series of childhood measles in London after World-War II (between 1945 and 1990), the vaccination program (starting in 1968) changed dramatically the number of cases \cite{Keeling2001,Earn2000,Becker2019}. Beside the medical treatment, if available, other artificially forced decline in contact rate (such as gradual lockdown prescribed by administration) can be imitated by Eq.~(\ref{eq:b0t}) resulting in a changing epidemic. This secular variation of parameter $b_0$ gives a clear analogy to climate change models and the framework applied in this field.

\section{\label{sec:stat_epidemic} Stationary epidemic}

\subsection{\label{sec:parameter_ensemble}Parameter dependence}

\begin{figure}
\includegraphics[width=0.9\columnwidth]{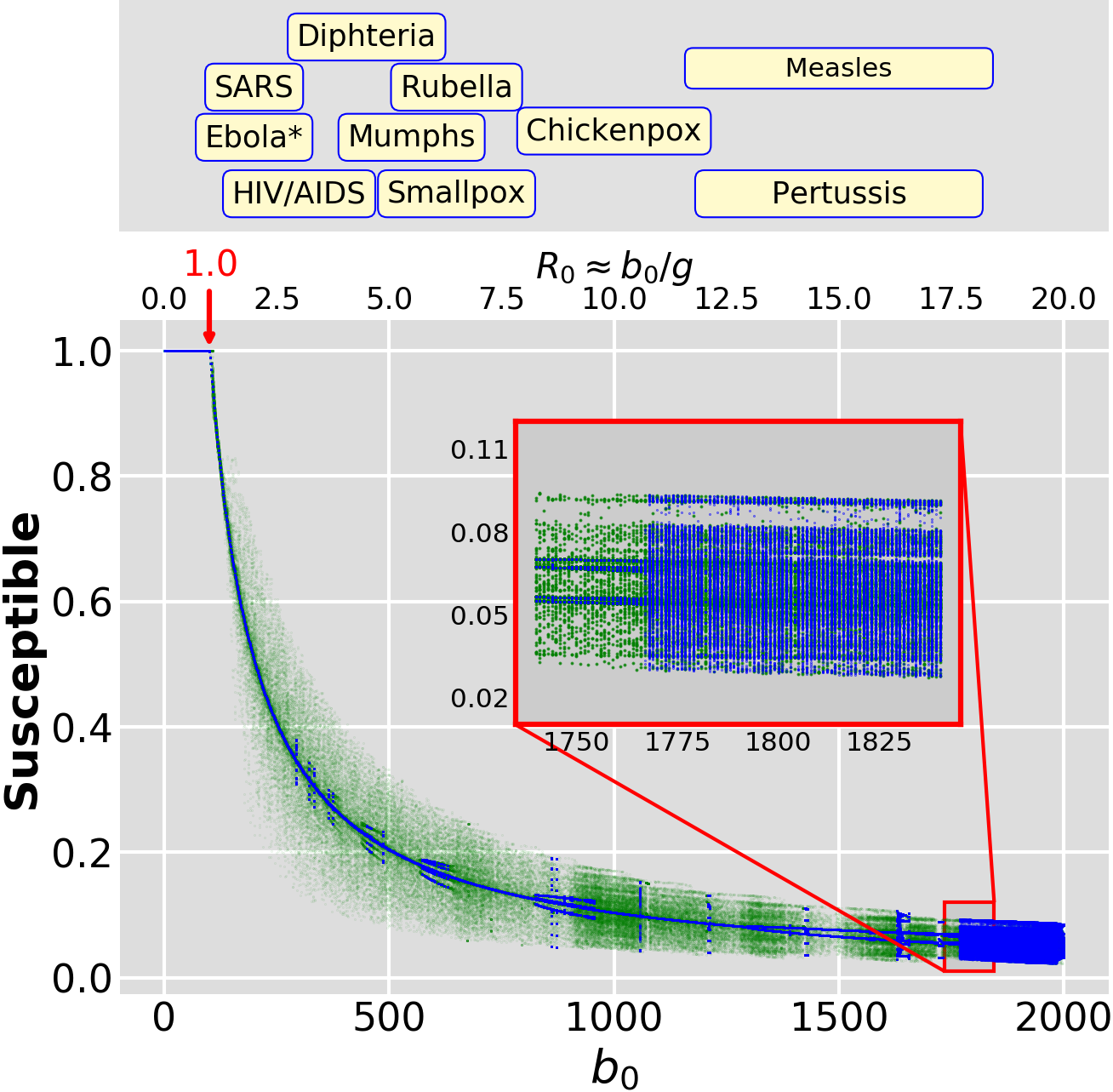}%
\caption{\label{fig:bif_diagram} Bifurcation diagram of the SEIR model with stationary $b_0$. The variable $S$ is plotted on the same year of the day starting at $t_0=0.$ The classical diagram (blue curve) is obtained by integrating 2000 initial conditions for $t=15000$yrs and the last 500 points stored. For low parameter values cycles are determined by annual periodicity. There are many period doubling visible (shorter blue segments), however, the first one that routes to chaos starts at $b_0\approx 1250.$ The end-points of the same trajectories are plotted after 500yr iteration indicating long-lived transients (green). Clearly finite time chaotic motion unfolds the extra structure of bifurcation diagram. Upper x-axis evaluates the basic reproductive ratio derived from system parameters. $R_0=1$ is also marked (red arrow) to indicate the edge of endemic equilibrium. Long-term dynamics for $R_0<1$ ($b_0<100$) implies a dying out infection in population. The inset display real permanent chaos. Top box displays the $R_0$ of common diseases for comparison only. (Ebola$^*$: based on 2014 Ebola outbreak.)}%
\end{figure}

The bifurcation diagram basically reveals the long-term dynamics, excluded the initial transients, in a given parameter range. Classically, one single trajectory sampled by stroboscopic map serves the blue shape of bifurcation diagram in Figure~\ref{fig:bif_diagram}. In case of the stationary or \textit{stationary} epidemic ($b_0=$ const.) complex dynamics arises for certain parameter values in the SEIR model, $b_0 \gtrsim 1770$. We should note, and it is going to be essential in our analysis, that the bifurcation diagram can also be established in a different way. That is, a large number of initial conditions are placed in the phase space and their evolution is monitored for sufficiently long time,
but much shorter (one or two orders of magnitude) than that of the single trajectory considered above. Due to the ergodicity of the stationary chaotic dynamics the end-points of the ensemble members portray exactly the same pattern \cite{Kaszas2016}. 

\begin{figure}
\includegraphics[width=0.9\columnwidth]{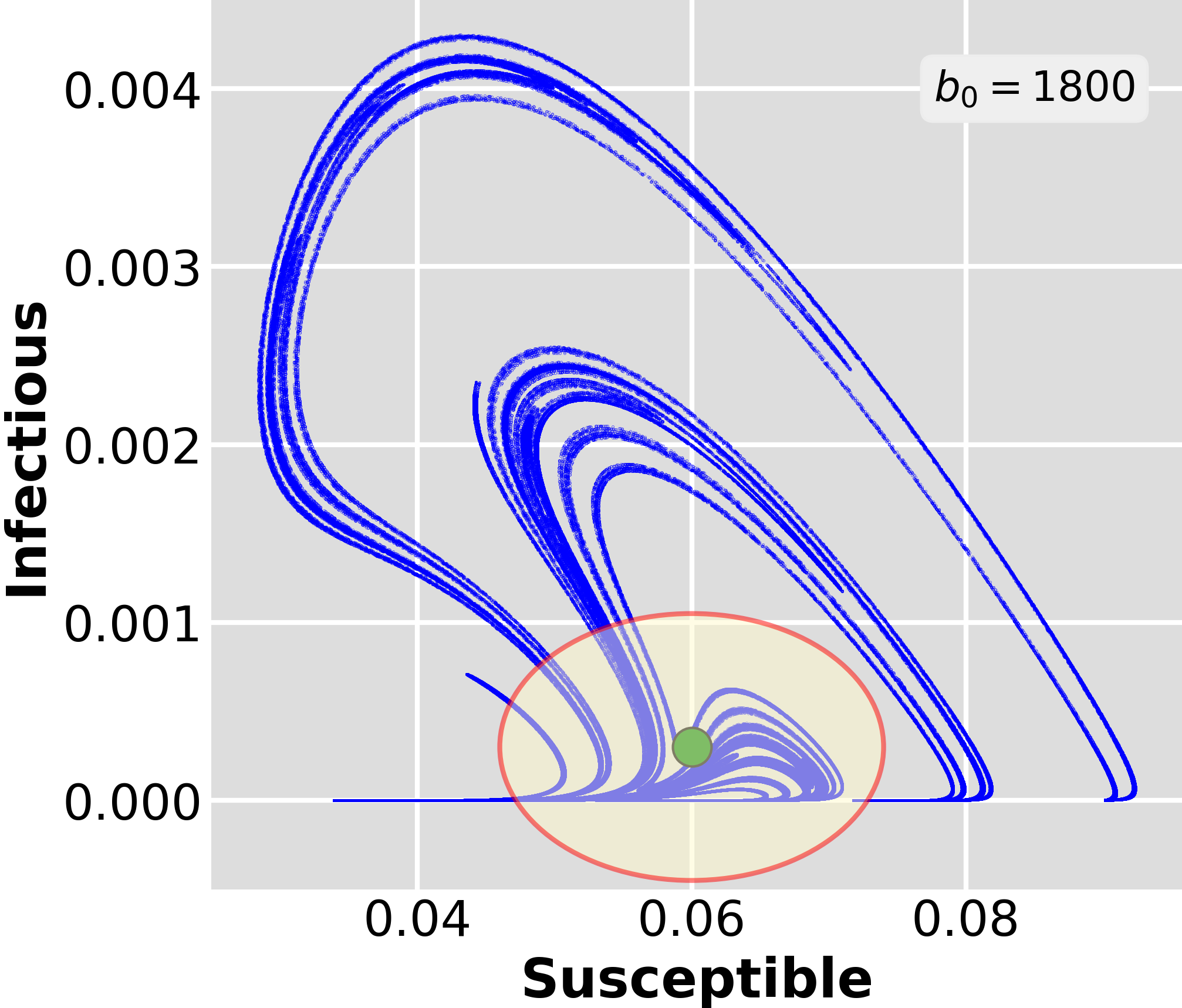}%
\caption{\label{fig:phase_space}The stationary chaotic attractor in the $S-I$ phase plane. The ensemble is plotted at three different time instants: $t=500,\;1250,\;2500$ years with perfect overlap. The green circle depicts the $\langle S\rangle$ and $\langle I\rangle$ values from Fig.~\ref{fig:transient} (middle panel) while the ellipse indicates the associated standard deviations (bottom panel), respectively. Note that the green circle is a weighted average (a kind of 'barycentre') of the trajectories along the attractor rather than its geometrical focus.}%
\end{figure}

The same applies for the phase space pattern, too. Fig.~\ref{fig:phase_space} depicts the chaotic attractor obtained after $T=2500$yr integration and 1600 initial conditions distributed uniformly in phase space. The same picture is obtained by one single trajectory after 200k year simulation.  
A representative $S-I$ phase portrait is pictured in Figure~\ref{fig:phase_space}. 
The chaotic attractor ($b_0=1800$) has been established fairly early and it remains unchanged when viewed after integer multiples of $T=$ 1yr. Nevertheless, choosing a different day of the year, the same applies eith a different pattern in $S-I$ plane.

Transient chaos, complex behavior on finite-time scales \cite{Lai2011}, manifests before a trajectory comes to an attractor, both in dissipative as well as in conservative systems. In general, the attractor can be a simple object in the phase space, for instance, a fixed-point or a limit cycle. Following the time evolution of the ensemble before it approaches the attractor, one can capture other ingredients of the bifurcation diagram by considering the initial transients. To do this, the uniformly distributed 2000 trajectories are integrated for 500 years and the corresponding state space positions are stored. This feature becomes visible in the green vague domain that extends along the blue curve. Clearly, transient chaos has a significant contribution to the dynamics for $b_0 \gtrsim 125.$

\subsection{\label{subsec:ens_view}Ensemble view}

The initial transients are usually thrown away in long-term dynamical analysis. However, the complexity might appear just in this phase of motion as indicated by the bifurcation diagram in Figure~\ref{fig:bif_diagram}, already in the stationary case. Illustrating the role of finite time chaotic behavior in the SEIR epidemic model we track the evolution of 1600 different initial conditions distributed uniformly in a cube $S=[0.9999;1],$ $E=[0;0.00005],$ $I=[0;0.00005],$. 

\begin{figure}
\includegraphics[width=0.9\columnwidth]{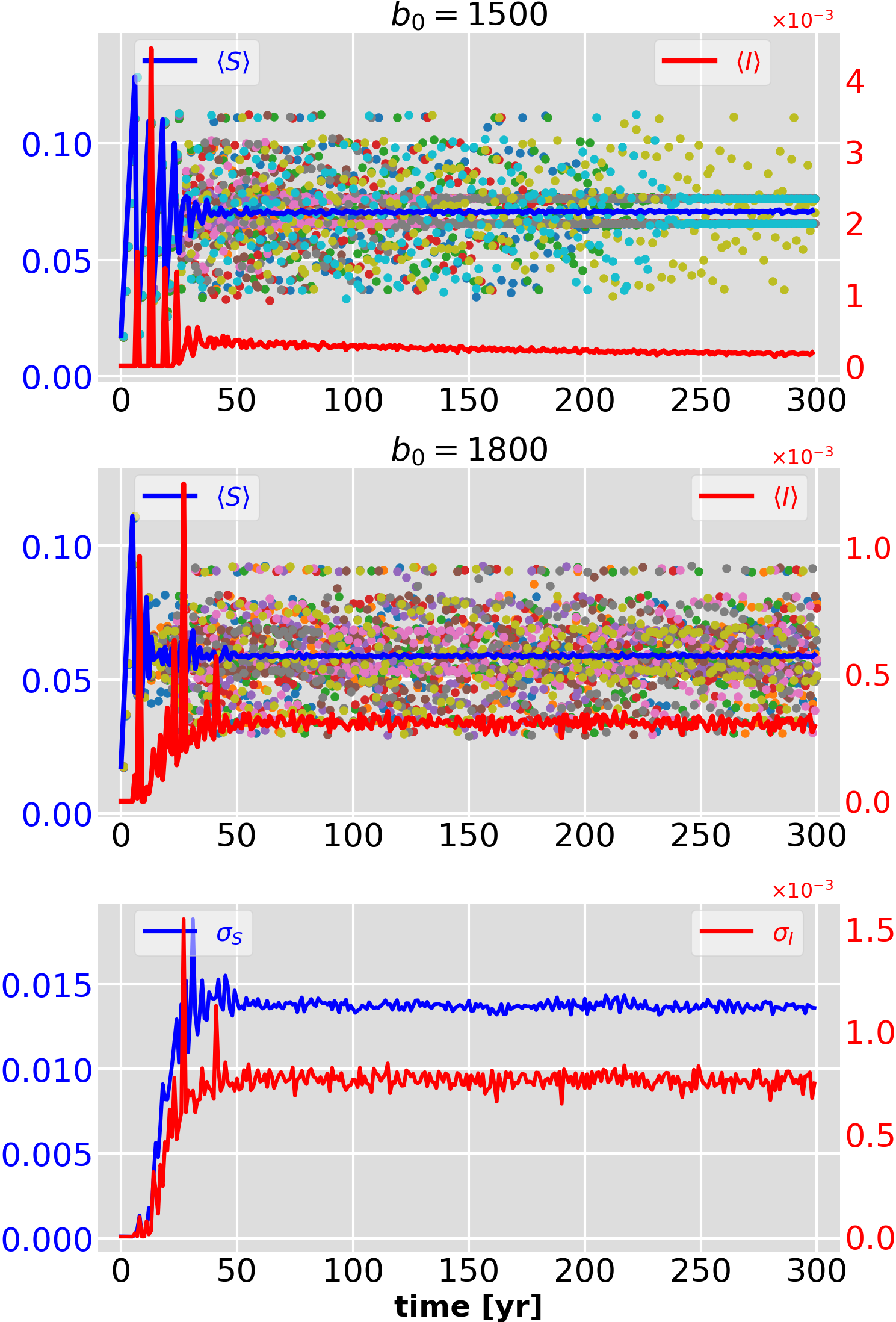}%
\caption{\label{fig:transient}Ensemble view of stationary dynamics. Top and middle: the mean of $S$ (blue) and $I$ (red) variables vs. time. Some of the individual trajectories are also shown sampled by stroboscopic map (only the $S$ co-ordinate). The individual trajectories arrives sooner or later the biannual fixed-point attractor (see for example the cyan parallel dots for $t>230$ in top panel). Bottom: standard deviations of the same variables as in the middle for $b_0=1800$. The constant value of average and standard deviation after the convergence time demonstrates that the ensemble has reached the attractor and then spread along it.}%
\end{figure}

Fig.~\ref{fig:transient} demonstrates how the individual members of the ensemble reach the attractor at different time instants. For those values of contact parameter when the epidemic shows biannual ($b_0=1500$) cycles one can observe clearly that the chaotic transients die out sooner or later (top panel). While in the case of permanent chaos ($b_0=1800$) it is not so, since the irregular property is perpetual (middle panel).

Introducing classical statistical measures on the ensemble one can quantitatively keep track of transient effect. The mean $\langle A(t) \rangle=1/N\sum_{i=1}^{N} A_i(t)$ (where $A_i(t)$ denotes an observable corresponding to the \textit{i}th member at time t) designates how fast the initial irregularity terminates and also adverts the average transient time. One can observe the similar characteristics for both the $\langle S \rangle$ and $\langle I \rangle$ curves, fixed-point and chaotic attractors (top and middle panels), respectively. 

In the stationary epidemic model one might expect that after some time the extent of the attractor remains constant suggesting that all of the individual trajectories in the ensemble have reached it. The standard deviation, 
\begin{equation}
    \sigma_A= (\langle A^2 \rangle-\langle A \rangle^2)^{1/2},
    \label{eq:sigma}
\end{equation}  
refers to the size of the attractor in the direction of $A.$  Figure~\ref{fig:transient} (bottom panel) depicts $\sigma_S$ and $\sigma_I$ in case of $b_0=1800.$ The initial small values of standard deviations show that the ensemble moves together at the beginning of integration. Then, the members spread out in the phase space ($\approx 30-50$ years) and after a while start to approach the attractor. This time is longer for parameters $b_0=1500$ (not shown) but $\sigma_S$ is smaller. 

Stationary epidemic with constant $\sigma_A$ essentially means that after the transients the phase space structure does not change in time. 
This can be visualized by using of ensemble approach. Without loss of generality we always start our simulation at $t=0$ (that corresponds to that part of the year with highest seasonal amplitude $b(t)$) and take the next picture about the ensemble at $t\mod T=0$ that coincides the same day of the year. 

\begin{figure}
\includegraphics[width=0.9\columnwidth]{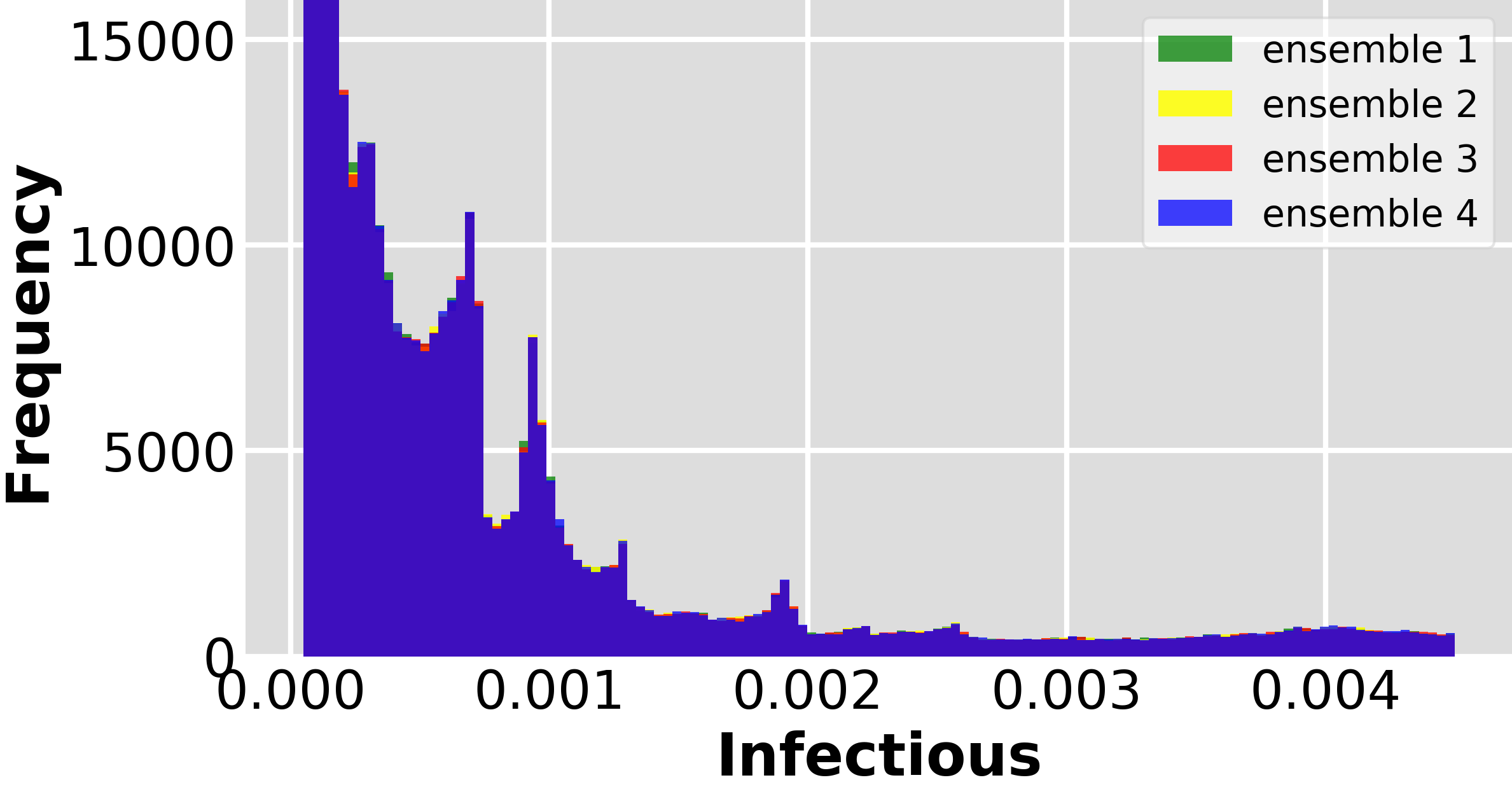}%
\caption{\label{fig:nat_dist}Natural distribution on the chaotic attractor ($b_0=1800$) projected onto the $I$ variable after $t=8500$ years. The distribution does not depend on the individual members of the ensemble. The frequency has been cut at 3500 for the clarity.}%
\end{figure}

The \textit{natural distribution} of a chaotic attractor \cite{OttBook} comes from the fact that the dots visit certain parts of the filamentary structure more frequently than others. Moreover, this distribution is stationary and does not depend on initial conditions. Instead of investigating a 3D frequency diagram (see supplementary Figure~S2), we plot the projection $I$ of trajectories wandering on the chaotic attractor in Figure~\ref{fig:phase_space}. To obtain Figure~\ref{fig:nat_dist}, a fine grid is defined in the $S-I$ phase plane and the number of points in each cell is plotted against the variable $I.$ The histogram consists of four different ensembles. Ensemble 1,2, and 3 covers the same volume in the phase space but involves slightly different initial state vectors. Ensemble 4 lays out initial conditions from other part of phase space. The distribution clearly shows the same pattern for all four ensembles.

\section{\label{sec:changing_epidemic}Epidemic change}

In a changing epidemic the parameter(s) of the system is(are) varying as time goes on. One might think that, say, a decreasing of the contact parameter $b_0$ means to walk along the bifurcation diagram slowly from right to left, and terminating at a safe destination with $R_0<1.$ In this section we will show that this idea is fairly naive due to the internal variability and the transient effects in the epidemic. To capture the dynamics properly in this scenario, the ensemble approach and the concept of snapshot attractor is desirable. 
\subsection{\label{subsec:snapshot_attr}Snapshot attractor geometry and natural distribution}

According to the common measles parameters we start the switch-off process, called also epidemic change, from the chaotic attractor ($b_0=1800$) according to Eq.~(\ref{eq:b0t}). 

\begin{figure}
\includegraphics[width=0.9\columnwidth]{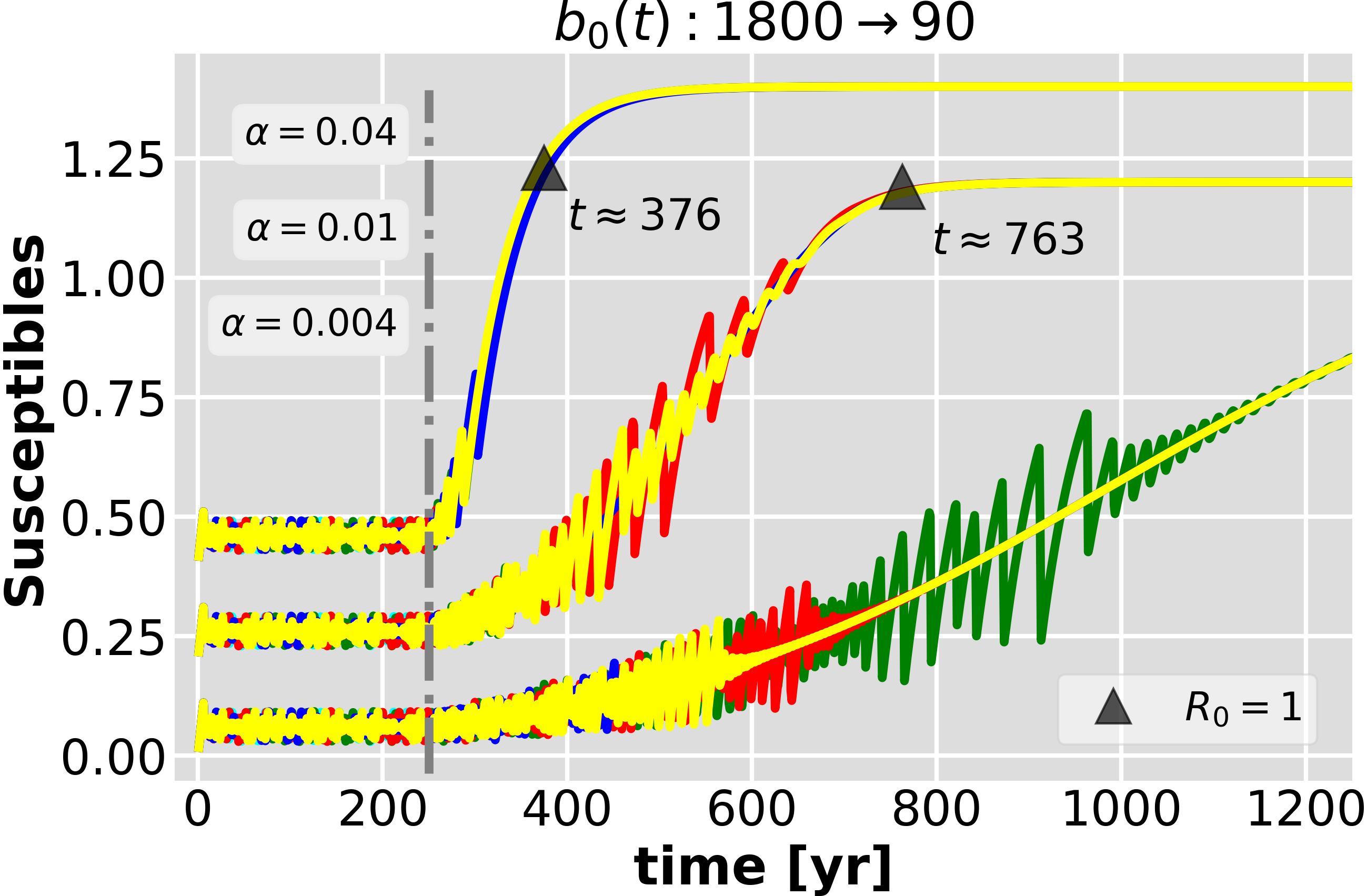}%
\caption{\label{fig:diff_traj}Four trajectories plotted in blue, green, red, and yellow wandering on the chaotic attractor and their fate (in $S$ variable) after parameter change sets in at $t=250$ yrs (vertical line). The two scenarios $\alpha=0.01,\;0.04$ are shifted by 0.2 units, respectively, for better visualization. The very peculiar time evolution of individual trajectories for different $\alpha$s suggests to use ensembles.}%
\end{figure}

\begin{figure*}
\includegraphics[width=0.9\textwidth]{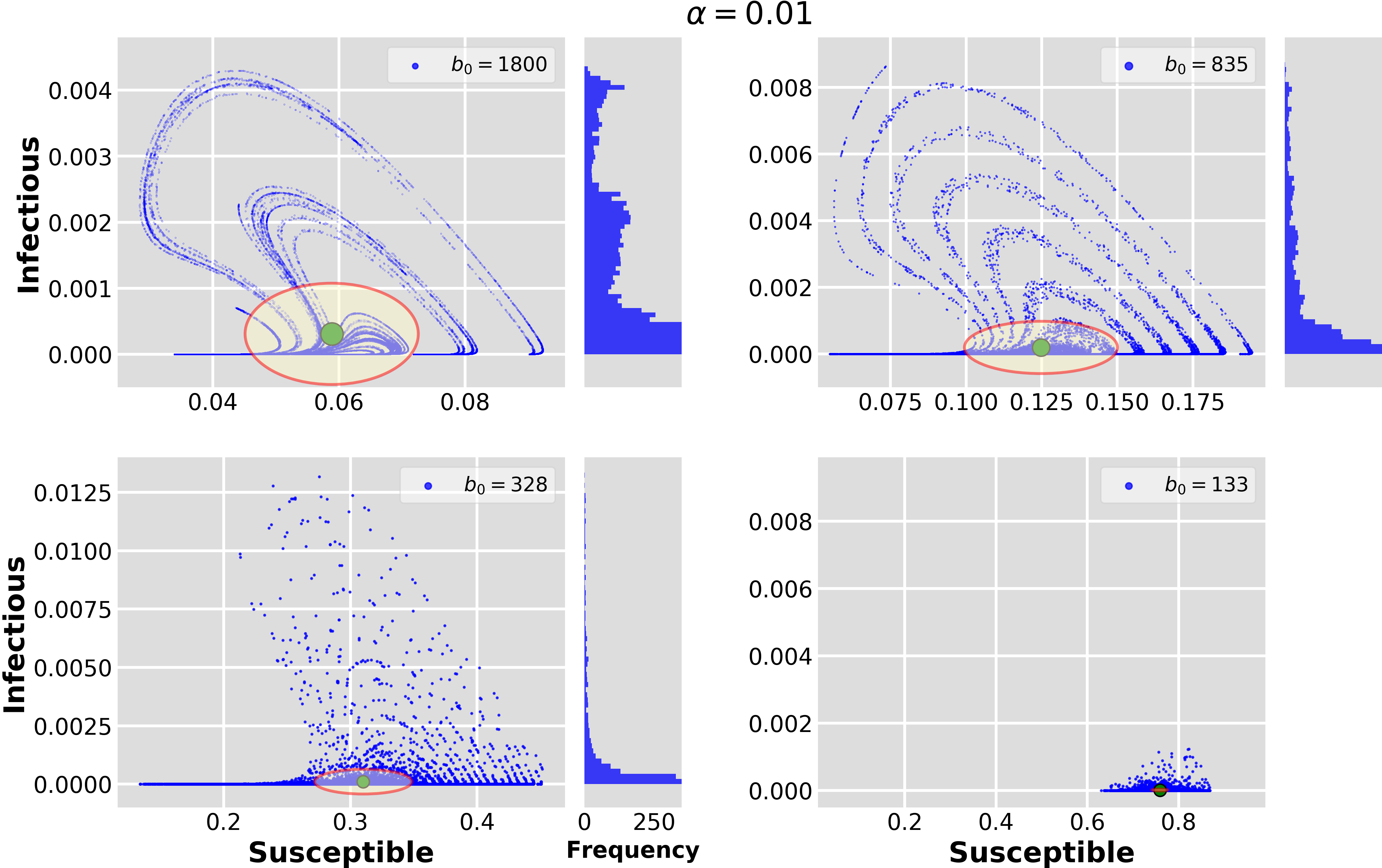}%
\caption{\label{fig:nat_dist_var} The evolution of the snapshot attractor with $\alpha=0.01.$ From left to right and top to bottom the ensemble is taken at $t=219,\;333,\;447,\;618$ yrs. The first picture still corresponds to stationary dynamics (before $t_{\mathrm{st}}=250$ yr) with the original $b_0=1800.$ Filamentary pattern persist for lower $b_0$ values, too, where no stationary chaotic attractor exists in phase space according to the bifurcation diagram (Fig.~\ref{fig:bif_diagram}). Still to be noted the physical extension of the snapshot attractor that is also time-dependent and their size varies significantly in both directions. Compare the scale of axes in different panels. The green circles and the red ellipses denote the same quantities as in Figure~\ref{fig:phase_space}. Slim panels to the right of the $S-I$ phase planes display the natural distribution projected onto the $I$ variables. For $b_0=133$ most of the phase points accumulate around the fixed-point attractor ($I\approx 10^{-3}$), therefore, no histogram is presented. Initial conditions cover the following state space volume: $S=[0.9999;1],$ $E=[0;0.00005],$ $I=[0;0.00005].$}%
\end{figure*}

Recent studies \cite{Kaszas2016} reveal that starting from the chaotic attractor well after that the trajectories reached it, the evolution of different states might be rather diverse. That is, transient dynamics becomes important again while the parameter change appears in the system. In Figure~\ref{fig:diff_traj}, four trajectories are selected from the chaotic attractor and their time dependence is followed. It can be immediately seen that different colors reach the fixed-point attractor at different times. For instance, in case of $\alpha=0.004$ (bottom curve), first the blue, then the yellow, red, and green curves arrive at the fixed-point attractor. One can pick up other trajectories that will have longer or shorter oscillations. Furthermore, the previous order of colors, i.e. transient times, might change with $\alpha,$ as demonstrated by the middle curve. Therefore, we can point out, that it is worth investigating several trajectories simultaneously instead of following individuals. However, after the parameter shift is switched on, the shape of the chaotic attractor starts to vary and forms a time-dependent set called \textit{snapshot attractor.} Thus, snapshot attractor is an object in the phase plane that contains the whole ensemble at a given time instant. We note that this mutation is not a result of the particular choice of initial phase (day of the year) of the mapping rather than the change of parameter. The geometrical alteration of the snapshot attractor is followed by the change of distribution on it. Nevertheless, its form and the distribution is independent of the choice of the ensemble and the onset ($t_{\mathrm{st}}$) of parameter decay.

Figure~\ref{fig:nat_dist_var} exhibits snapshot attractors drawn by $N=10^4$ initial conditions during the epidemic change. The mean contact rate decreases from top to bottom and left to right as specified on each panel. Thus, every single plot corresponds to a different time instant of simulation. The top left panel, taken at $T=219$yr, coincides with the attractor in Figure~\ref{fig:phase_space} since the parameters of the system are the same for $t<t_{\mathrm{st}}=250$ yr. A more interesting aspect shows up on the following two panels. A filamentary structure, the fingerprint of chaotic behavior, still dominates the pattern although the parameter $b_0$ is well below of its original value (1800) as well as  of 1770 where large extended chaotic attractors are formed in the  bifurcation diagram (Fig.~\ref{fig:bif_diagram} inset). In other words, for those contact rates ($b_0=835,\,328$) no chaotic behavior is anticipated in the stationary dynamics.

Although, Figure~\ref{fig:nat_dist_var} indicates the transmutation of the snapshot attractor for decay exponent $\alpha=0.01$ one can obtain similar alteration for different switch-off rates too. Slower scenario (e.g. $\alpha=0.004$ in Eq.~(\ref{eq:b0t})) allows the attractor to keep its filamentary shape and maintain chaotic dynamics longer. From other perspective, the same parameter value $b_0$ is reached later while $\alpha$ is smaller. The opposite is true for a faster parameter change, say $\alpha=0.04.$ It can also be shown that approximately the same pattern belongs to the same contact value regardless the rate of change. That is, the faster the contact rate decay, the less pronounced the complexity in phase space. 

Other interesting feature is that although the contact rate $b_0$ decreases monotonically, the size of the snapshot attractors might increase, Figure~\ref{fig:nat_dist_var} top right panel. This fact yields that the domain accessible by the dynamics can be larger. Thus the possible ($S,I$) pairs extend to larger domain in phase space even for smaller contact parameter. Furthermore, the average (green circles) and the standard deviation (red ellipses) also change in time. And this temporal behaviour cannot obtained from the classical view, only by using snapshot attractors.

\begin{figure}
\includegraphics[width=0.9\columnwidth]{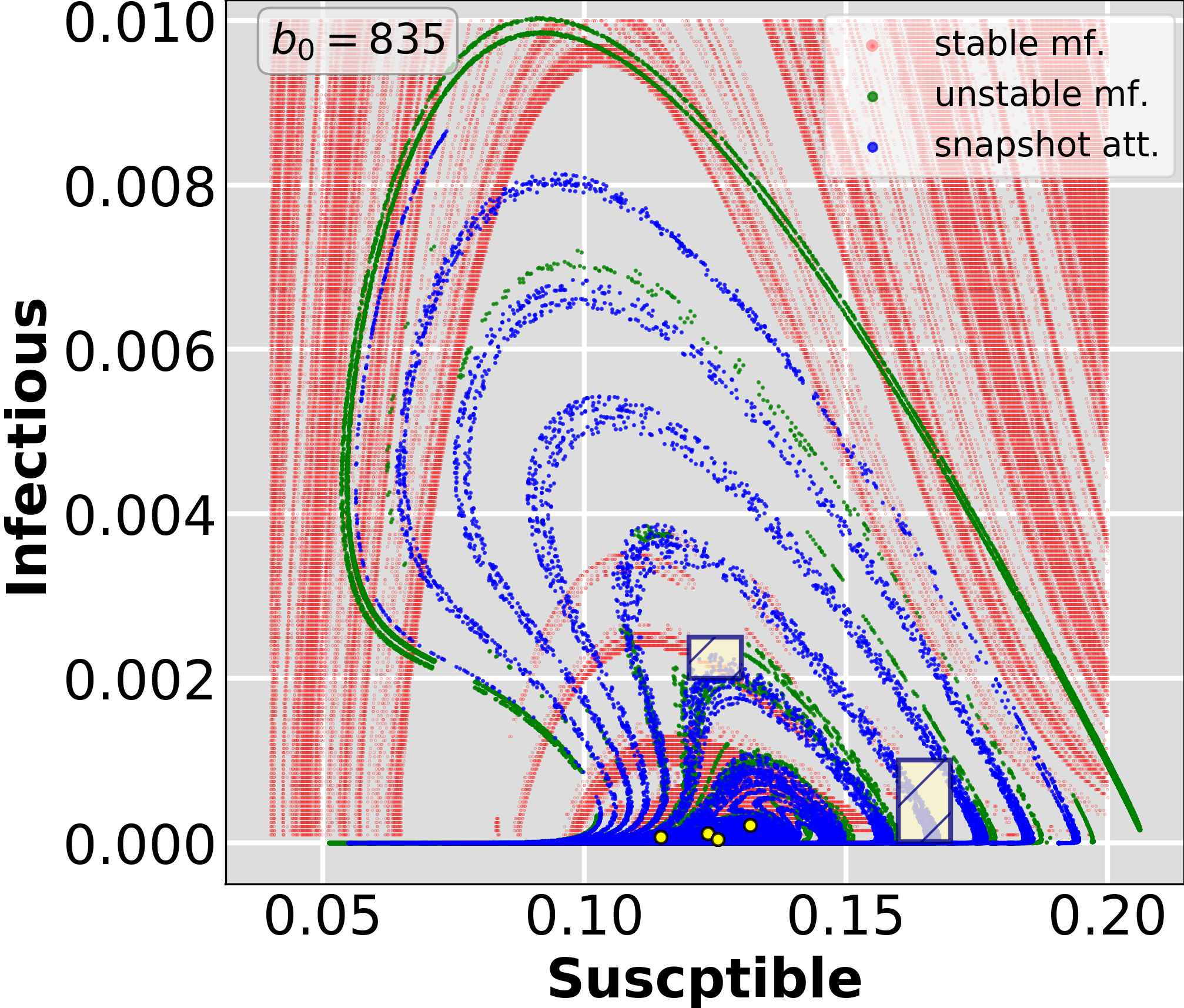}%
\caption{\label{fig:saddle_manifolds} The snapshot attractor (blue) and stable/unstable (red/green) manifolds of the stationary chaotic saddle at $b_0=835.$ Discontinuities along the unstable manifold reflect to the escape conditions of the numerical scheme. The long-term dynamics of the stationary model is visualized by the four tiny yellow circles around $(S,I)\approx(0.125,0.0004)$ as fixed-point attractors. The snapshot attractor coincides with the top right phase portrait in Figure~\ref{fig:nat_dist_var}}%
\end{figure}

To understand this phenomenon we should recall the concept of \textit{chaotic saddle.} This non-attracting set with its stable and unstable manifolds in phase space is responsible for the finite time chaotic behavior \cite{Lai2011}. To construct the saddle numerically, we define two holes in the phase space (black rectangles in Figure~\ref{fig:saddle_manifolds}). Then, a large number of initial conditions ($N=3\cdot 10^5$) are distributed uniformly in the region $S=[0.04;0.2],$ $E=[0;0.00005],$ $I=[0.00001;0.01]$ and the trajectories a integrated for $T=30$ years. Only those trajectories are kept that do not enter the holes during the integration time. The initial conditions belonging to these trajectories draw the stable manifold of the saddle (red dots in Fig. ~\ref{fig:saddle_manifolds}), while those that are just before leave plot the unstable manifold (green dots.) The saddle itself is the intersection of its manifolds, not shown here. The average lifetime of chaos (the inverse of escape rate, $\kappa$) can be estimated by calculating the time distribution of the non-escaped trajectories.   

It is also known from the theory of transient chaos that the saddle's manifolds have filamentary design just like the snapshot attractors in Fig.~\ref{fig:nat_dist_var}. Analogously to chaotic attractor, stationary dynamics with constant driving amplitude defines a stationary chaotic saddle related to certain parameter values of the system. Due to the continuous adjustment of the contact rate, $b(t),$ in the changing epidemic model, the trajectrories do not have time to reach the attractor belonging to a given $b_0$ value. Consequently, a time dependent chaotic saddle is considered, whose unstable manifold approximates the snapshot attractor \cite{Kaszas2016}. This stationary saddle persist for very low $b_0$ values and its unstable manifold controls the finite time complex epidemic dynamics, Figure~\ref{fig:saddle_manifolds}. 

We emphasize at this point that the natural distribution associated to the snapshot attractor also changes in time following the geometrical reorganization of the phase space pattern (histogram visualization in Fig.~\ref{fig:nat_dist_var}).

\subsection{\label{parallel_real}Parallel epidemic realizations}

Similarly to climate research we can define the concept of \textit{parallel epidemic realizations.} In standard disease models like Eq.~(\ref{eq:seir}) this can be done by using a large number of initial conditions as an ensemble in phase space and following their evolution as discussed in Section~\ref{sec:parameter_ensemble}. This picture can be imagined as many copies of the epidemics obeying the same physical laws and being affected by the same time-dependent forcing\cite{Tel2019}.

As we have seen before, in case of a changing epidemic, the time-dependent forcing has an impact on the natural measure of the snapshot attractors. This fact results in a temporal change of the average values as well as internal variability. To quantify the internal variability of the system (\ref{eq:seir}) statistical measures over the ensemble should be proposed like in the case of stationary epidemic. The variance or the standard deviation $\sigma_A$ in Eq.~ (\ref{eq:sigma}) of the ensemble characterize the fluctuations around the averages indicating the inherent internal variability. 

\begin{figure*}
\includegraphics[width=0.9\textwidth]{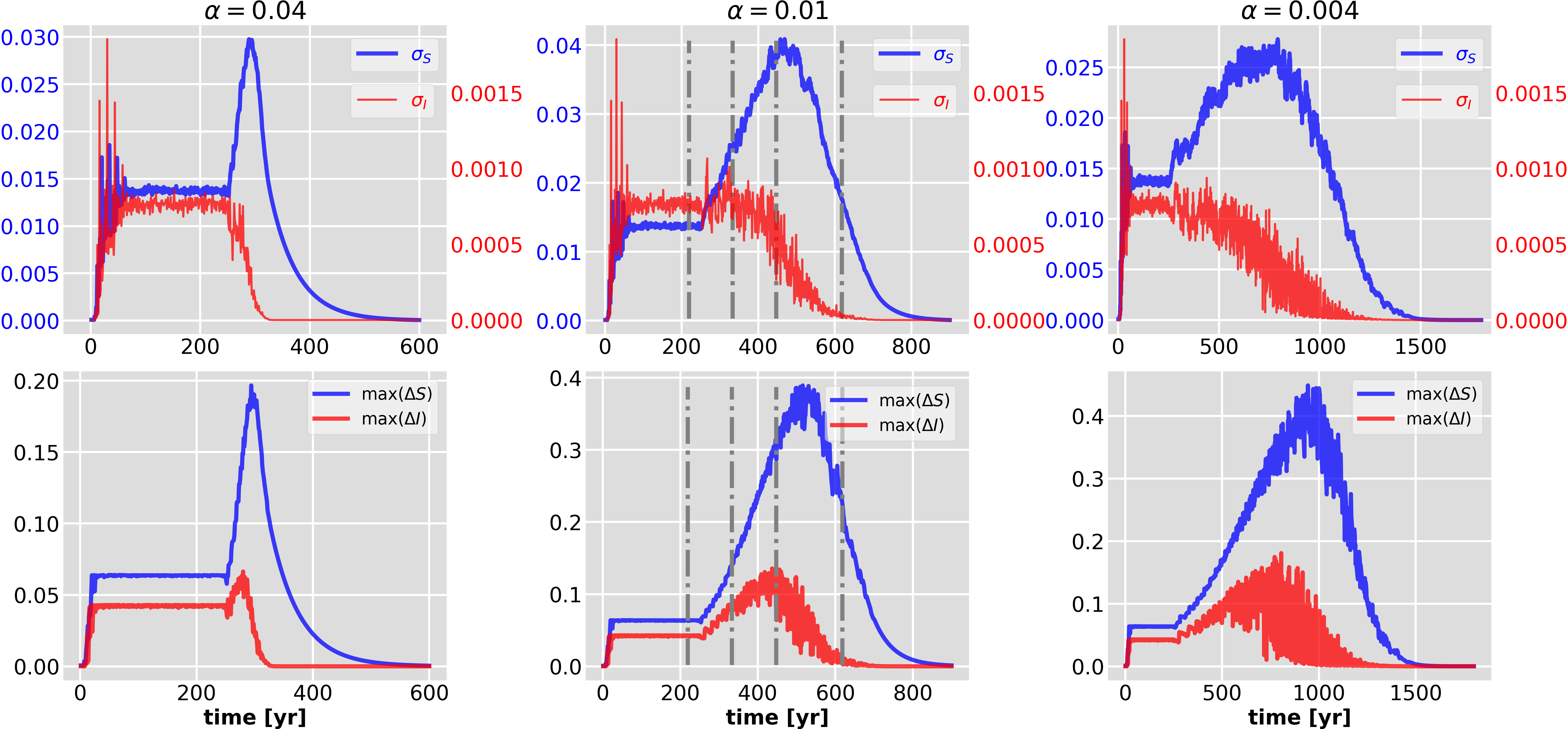}%
\caption{\label{fig:statistics} Statistical measures of SEIR-ensembles. Top: The standard deviation of variable $S$ and $I$ for various decay processes. The epidemic change starts at $t_{\mathrm{st}}=250$ yrs well after the convergence time ($t_c\approx 50$ yrs). Vertical dashed lines mark the time instants corresponding to phase space portraits in Figure~\ref{fig:nat_dist_var}. Bottom: The linear size of snapshot attractors in both variables indicate the role of transient chaos and internal variability of dynamics depending on the decay rate $\alpha.$}%
\end{figure*}

Top panels of Figure~\ref{fig:statistics} show the ensemble standard deviation $\sigma_S$ (blue) and $\sigma_I$ for different decay rates $\alpha.$ The parallel epidemic realizations contain 1600 numerically integrated trajectories sampled at integer multiplies of years. The convergence time $t_c,$ the time needed of the ensemble to reach and spread along the attractor, is found to be $t_c=50$yr. Both $\sigma_S$ and $\sigma_I$ are constant before $t_{\mathrm{st}}=250$yr since till then the seasonal driving is constant, i.e. $b_0=1800$yr$^{-1}.$ Right after this the contact rate starts to decrease ($t>250$yr) according to Eq.~(\ref{eq:b0t}) both the mean state and the internal variability of the epidemic changes with time.  

In changing epidemic, first, the graph of $\sigma_S(t)$ increases, after the maximum the trend follows nearly $b_0(t),$ it decays to zero illustrating that the size of the attractor shrinks and asymptotically reaches the neighbourhood of a regular fixed-point attractor, $(S,I)=(1,0),$ as expected for $b_0\approx 90$ ($R_0=0.9$). The larger the $\alpha,$ the more regular, i.e less filamentary, the phase space geometry at the same time, see bottom panels in Fig.~\ref{fig:statistics}. 

The shape of the standard deviation can be explained by transient chaos that occurs during the parameter change. The size of the green shaded band around the main feature in the bifurcation diagram (Fig.~\ref{fig:bif_diagram}) already indicates that the size of the phase space region filled with transient chaos increases as $b_0$ reduces. To demonstrate this we call the attention to the horizontal dimension of the snapshot attractors in Figure~\ref{fig:nat_dist_var}. The smoothly decreasing profile of the standard deviation after reaching the maximum refers to the existence of transient effect up to very small parameter values.

Different maximum values of the $\sigma_S(t)$ curves indicate a non-trivial relation between the internal variability and changing rate of $b_0.$ The largest value corresponds to $\alpha=0.01$ (top middle panel), the other two scenarios ($\alpha=0.04$ and 0.004) show roughly the same maximum size of the snapshot attractor, albeit, at different time instants. Bottom panels present the maximal physical extension of the snapshot attractors versus time. These plots also supports our observations that the size of the attractor first increases and the shrinks to be a fixed-point attractor. 

One possible explanation of this property is the relative ''coupling'' between the timescales, that is, decay of forcing defined by the exponent $\alpha$ and the average lifetime of chaos in stationary dynamics at specific parameter $b_0.$ From a geometrical point of view, how close the snapshot attractor evolves to the unstable manifold of the stationary chaotic saddle at particular contact parameter. A detailed exploration of this feature is, however, postponed to a future study.

\section{\label{sec:concl}Final remarks}

Parallel climate realizations, an effective and new framework in climate research, has been adopted to an  epidemic model to explore the fading of the complexity due to the systematic switch off the driving mechanism. The mathematical concept of snapshot attractors and their natural distribution demonstrates that single time series analysis is not capable to reflect the complex dynamics of a changing epidemic. Instead of monitoring isolated events, the ensemble view of trajectories -- parallel epidemic realizations -- and its statistical description is desirable. 

The temporal change of the attractor geometry and the distribution on it reveals the internal variability of the dynamics.  

The extension of the bifurcation diagram indicates the importance of transient chaos in the stationary dynamics as well as during the continuous parameter shift. No matter whether we start from a stationary state of the long-term dynamics, the switch-off process activates the hidden parts of the bifurcation diagram. Thus, the underlying non-attracting object, the chaotic saddle, or more precisely its unstable manifold, organizes the system's evolution. 

The dissipative relaxation timescale, the inverse of the phase space contraction rate based on the divergence of the vector field of system (\ref{eq:seir}), is extremely low, $\sim 2-5$ days, for measles. The other characteristic times are, the inverse of the switch-off rate and the escape rate from the chaotic saddle are, for comparison, $\alpha^{-1}\approx 25-250$ yrs, $\kappa^{-1}\approx 30-35$ yrs, respectively. This latter is obtained a few particular contact rate. This implies that the switching off process, with parameter $\alpha$ used in this study, is not quasistatic \cite{Kaszas2016}. In other words, there is not enough time for trajectories to reach the stationary attractor, either it is chaotic or regular, due to the nonstop parameter variation. Therefore, the dynamics of the switching off process shows much more complexity than that of the stationary scenario.

In the spirit of the investigated SIR-like classical epidemic model (\ref{eq:seir}) with parameter shift \footnote{Generally speaking, the time variability of $b_0$ always results in a changing phase space structure according to the theory  of snapshot attractors.} of the contact rate $b(t)$ we emphasize the importance of the finite time irregular dynamics. Our results reinforce the ''hidden'' complex transients, and also the importance of time-dependent snapshot attractors not just for an epidemic with large reproductive ratio, such as  measles and chickenpox, but for those with lower values ($R_0\gtrsim 1 $) too.  

The main conclusion of this study sheds light on the importance of the ensemble view and parallel realizations in epidemic dynamics. Fig.~\ref{fig:nat_dist_var} illustrates the general features one expects in the (long-term) prediction of any epidemics. Predictions based on individual simulations are not reliable since, due to the chaotic nature of the dynamics, they can lead to many possible results, to any point of the snapshot attractor belonging to the time instant of the prediction. The ensemble approach is able to treat all possible outcomes as a whole, and predicts even the probability of the different permitted epidemic states. When convenient, statistical moments of this distribution can be determined, e.g. the average (providing the most typical epidemic outcome), or the variance (characterizing how broad the distribution of the permitted states is), but higher order moments might also be useful. Anyhow, one can thus follow how the statistical prediction changes in time.


\begin{acknowledgments}
We benefited from useful discussions with T. T\'el. This work was supported by the NKFIH Hungarian Grants K125171. The support of Bolyai Research Fellowship and \'UNKP-19-4 New National Excellence Program of Ministry for Innovation and Technology is also acknowledged. 
\end{acknowledgments}

\bibliography{refs_sir}

\section*{\label{suppl}Supplementary material}
\renewcommand{\thefigure}{S\arabic{figure}}
\setcounter{figure}{0}

\begin{figure}
\includegraphics[width=\columnwidth]{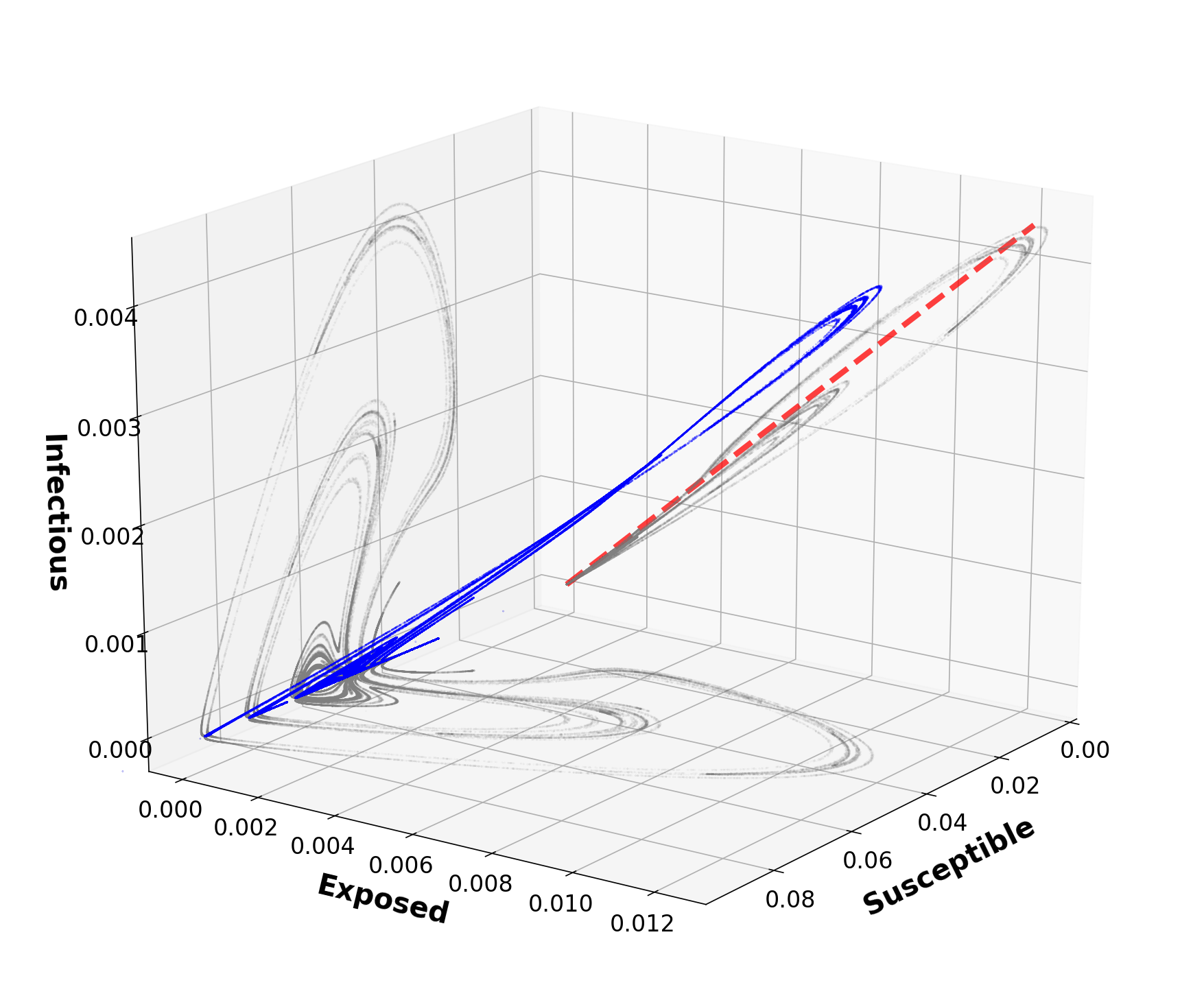}%
\caption{\label{fig:3D_phase_space} Three dimensional chaotic attractor in the stationary SEIR model. The parameters are defined in the main text. It is striking that the infectious ($I$) and exposed ($E$) classes are approximately linearly related \cite{Aron1984} (according to the ratio $g/a$). The relation is also indicated with red dashed line in the $I-E$ surface. Thus, one can investigate the $S-I$ projection without loss of generality. This has been done in the present work.}%
\end{figure}

\begin{figure}
\includegraphics[width=\columnwidth]{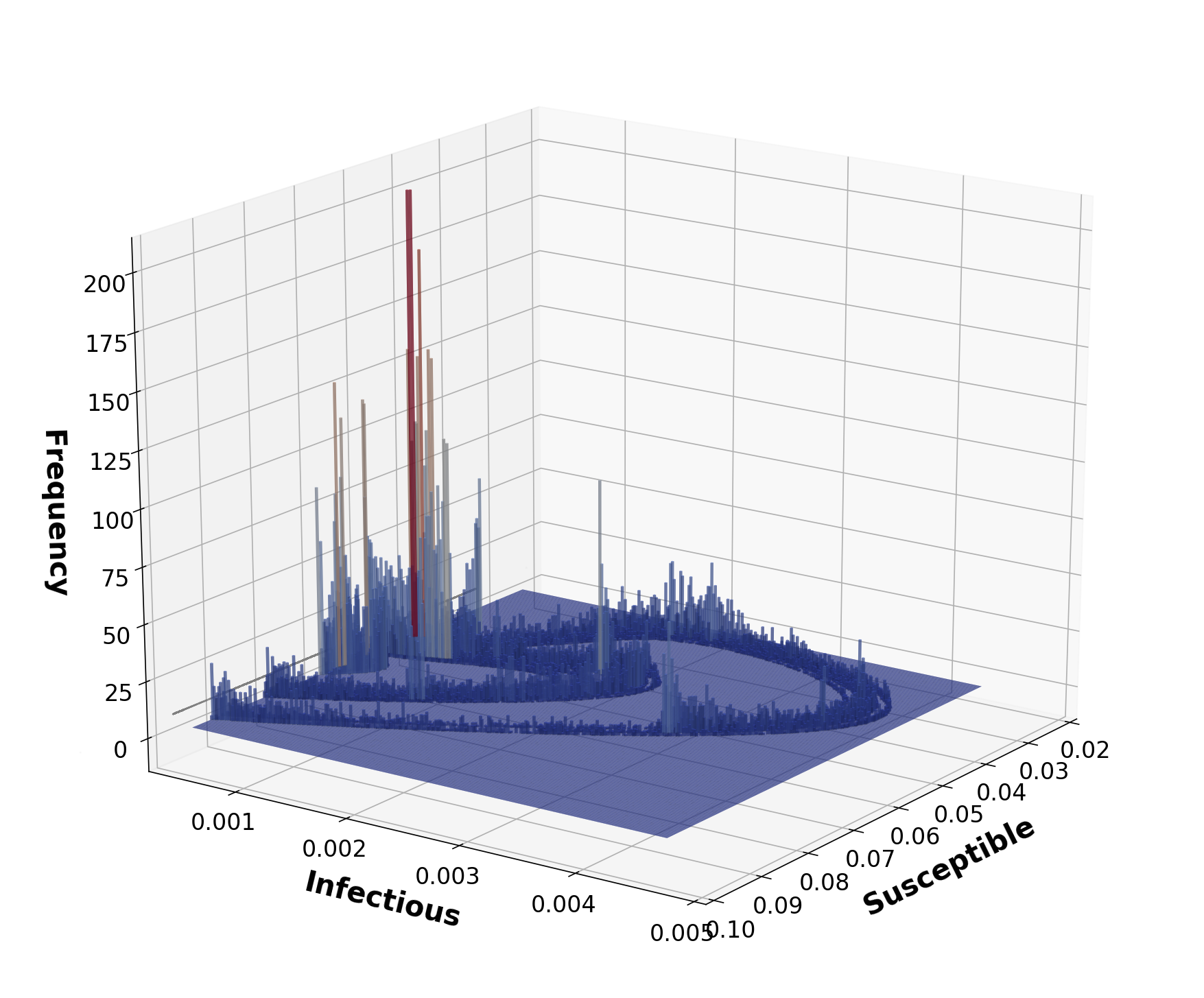}%
\caption{\label{fig:nat_dist_3D}Natural distribution on the chaotic attractor of Fig.~\ref{fig:3D_phase_space}. A fine grid is defined in the $S-I$ plane and the number of phase points in each cell is recorded. The histogram shows that some part of the attractor are more frequently visited than others assigning the natural measure on it. Note that the height of the bars along the $I\approx 0$ axis is 1-2 order of magnitude larger than in the rest of the plot.}%
\end{figure}

We present two illustrative plots related to the phase space structure of SEIR model (\ref{eq:seir}) that complement the overall picture of the paper but lack of the plots in the main text does not violates the understanding of the basic idea.

\end{document}